\documentclass[twocolumn,prc,aps,showpacs]{revtex4}
\usepackage{graphicx}

 \def\gsim{\mathrel{\rlap{\lower4pt\hbox{\hskip1pt$\sim$}}
 \raise1pt\hbox{$>$}}}

 \newcommand\beq{\begin{equation}}
 
 \newcommand\eeq{\end{equation}}
 \newcommand\beqn{\begin{eqnarray}}
 \newcommand\eeqn{\end{eqnarray}}

\def\fm{\,\mbox{fm}}
\def\GeV{\,\mbox{GeV}}

\def\lsim{\mathrel{\rlap{\lower4pt\hbox{\hskip1pt$\sim$}}
    \raise1pt\hbox{$<$}}}         
\def\gsim{\mathrel{\rlap{\lower4pt\hbox{\hskip1pt$\sim$}}
    \raise1pt\hbox{$>$}}}         

\def\fm{\,\mbox{fm}}
\def\GeV{\,\mbox{GeV}}

\def\s0{\sigma_0(s)}

\def\beq{\begin{equation}}
\def\eeq{\end{equation}}

\def\beqy{\begin{eqnarray}}
\def\eeqy{\end{eqnarray}}

\newcommand{\ber}{\begin{displaymath}}
\newcommand{\eer}{\end{displaymath}}
\newcommand{\bey}{\begin{eqnarray}}
\newcommand{\eey}{\end{eqnarray}}

\def\beq{\begin{equation}}
\def\eeq{\end{equation}}

\def\beqy{\begin{eqnarray}}
\def\eeqy{\end{eqnarray}}

\begin{document}

\title{\bf Why heavy and light quarks radiate energy with similar rates}

\vspace{1cm}

\author{B. Z. Kopeliovich}
\author{I. K. Potashnikova}
\author{Iv\'an Schmidt}
\affiliation{Departamento de F\'{\i}sica, Centro de Estudios
Subat\'omicos, Universidad T\'ecnica Federico Santa Mar\'{\i}a,
\\and\\
Centro Cient\'ifico-Tecnol\'ogico de Valpara\'iso,\\
Casilla 110-V, Valpara\'iso, Chile}
\begin{abstract}
\noindent The dead cone effect has been expected to reduce the magnitude of
energy loss and jet quenching for heavy flavors produced with large $p_T$ in heavy ion
collisions. On the contrary, data from RHIC for open charm
production demonstrate a flavor independent nuclear suppression. We
show that vacuum radiation of a highly virtual quark
produced at high $p_T$ with its color field stripped off,  develops a much wider
dead cone, which
screens the one related to the quark mass. Lacking the field, gluons cannot be radiated
within this cone until the color field is regenerated
and the quark virtuality cools down to the scale of the order of the quark
mass. However, this takes time longer than is essential for the
observed jet quenching. This is the reason why the light and charm
quark jets are quenched equally. Open beauty is expected to be suppressed much less
within the $p_T$ range studied so far.
\end{abstract}



\maketitle

Energy loss via gluon radiation induced by a dense medium
\cite{gw,bdmps} may be an important source of nuclear suppression,
called jet quenching, observed at RHIC \cite{phenix,star} for
hadrons produced with large $p_T$ in heavy ion collisions. Based on
the dead cone effect \cite{dead-cone} in the radiation of heavy
quarks, a reduced rate of energy loss induced by a medium was found
in \cite{yura}. This can be easily understood classically, since a
current does not radiate if its trajectory is straight. The source
of induced radiation is the wiggling of the charge trajectory due to
multiple collisions. A heavy particle wiggles in the medium much
less than a light one, since its transverse speed is $m_q/m_Q$ times
smaller at the same momentum transfer. A very heavy particle always
propagates straight.

Correspondingly, it has been expected that jet quenching for heavy
flavors, charm and beauty, should be weaker than for light hadrons
\cite{miklos1}. Surprisingly, measurements at RHIC did not confirm
this expectation, since charmed and light  hadrons turned out to be
similarly suppressed \cite{phenix-charm,star-charm}. This
controversy has not been settled so far \cite{leitch,dong}, in spite
of the attempts to improve the calculations of energy loss
\cite{miklos2}.

Here we draw the attention that the result of \cite{yura} should be
applied to heavy ion collisions with precautions, since the quark
which undergoes induced energy loss does not come from infinity, but
originates from a hard reaction and its color field is stripped off.
Therefore, at the initial state of hadronization the quark has no
field to radiate, except the gluons with very large transverse
momentum and short coherence time. The initial high virtuality of
the quark, $Q\sim p_T$, creates its own dead cone, which is much
wider than the one controlled by the quark mass. The quark mass does
not play a significant role until the quark virtuality drops down to
values of the order of the quark mass. Here we study the interplay
of these two phenomena.

High-$p_T$ scattering of partons leads to an intensive gluon
radiation in forward-backward directions, which is related to the
initial color field of the partons shaken off by the strong
acceleration caused by the hard  collision. The
Weitz\"acker-Williams gluons accompanying the parton do not survive
the collision and lose coherence up to transverse frequencies
$k\lsim p_T$. Therefore, the produced high-$p_T$ parton is lacking
this part of the field and starts regenerating itself via radiation
of a new cone of gluons, which are aligned along the new direction.
This process lasts a long time proportional to the jet energy
($E=p_T$), since the radiation time (or length) depends on the gluon
energy and transverse momentum $k$ (relative to the jet axis), \beq
l_c=\frac{2E}{M_{qg}^2-m_q^2}= \frac{2Ex(1-x)}{k^2+x^2\,m_q^2}.
\label{100} \eeq Here $E=p_T$ is is the jet energy; $x$ is the
fractional light-cone momentum of the radiated gluon;
$M_{qg}^2=m_q^2/(1-x)+k^2/x(1-x)$ is the invariant mass squared of
the quark and radiated gluon.

One can trace how much energy is radiated over the path length $L$ by the gluons which have lost coherence during this time interval,
 \beq
\Delta E(L) =
E\int\limits_{\Lambda^2}^{Q^2}
dk^2\int\limits_0^1 dx\,x\,
\frac{dn_g}{dx\,dk^2}
\Theta(L-l_c),
\label{120}
 \eeq
 where $Q\sim p_T$ is the initial quark virtuality; the infra-red cutoff is fixed at $\Lambda=0.2\GeV$.
 The radiation spectrum reads
 \beq
\frac{dn_g}{dx\,dk^2} =
\frac{2\alpha_s(k^2)}{3\pi\,x}\,
\frac{k^2[1+(1-x)^2]}{[k^2+x^2m_q^2]^2},
\label{140}
 \eeq
 where $\alpha_s(k^2)$ is the running QCD coupling, which is regularized at low scale by replacement $k^2\Rightarrow k^2+k_0^2$ with $k_0^2=0.5\GeV^2$.
 
In the case of heavy quark the $k$-distribution Eq.~(\ref{140}) peaks at $k^2\approx x^2\,m_q^2$, corresponding to the polar angle (in the small angle approximation) $\theta=k/xE=m_q/E$. This is the dead cone effect \cite{dead-cone,yura}.

The step function in Eq.~(\ref{120}) creates another dead cone: no
gluon can be radiated unless its transverse momentum is sufficiently
high, \beq k^2>\frac{2Ex(1-x)}{L}-x^2m_q^2. \label{160} \eeq This
bound relaxes with the rise of $L$ and reaches the magnitude
$k^2\sim x^2m_q^2$ characterizing the heavy quark dead cone at \beq
L_q= \frac{E(1-x)}{xm_q^2}. \label{180} \eeq We see that $L_q$ for
beauty is an order of magnitude shorter than for charm, but linearly
rises with the jet energy.

The characteristic length $L_q$ may be rather long, since gluons are radiated mainly with small $x$. For instance, for a charm quark with $E=p_T=10\GeV$ the sensitivity to the quark mass is restored at $L\gsim1/x\,\fm$. Only at longer distances, $L\gsim L_q$, the dead cone related to the heavy quark mass sets up, and the heavy and light quarks start radiating differently.

\begin{figure}[htb]
 \includegraphics[height=7cm]{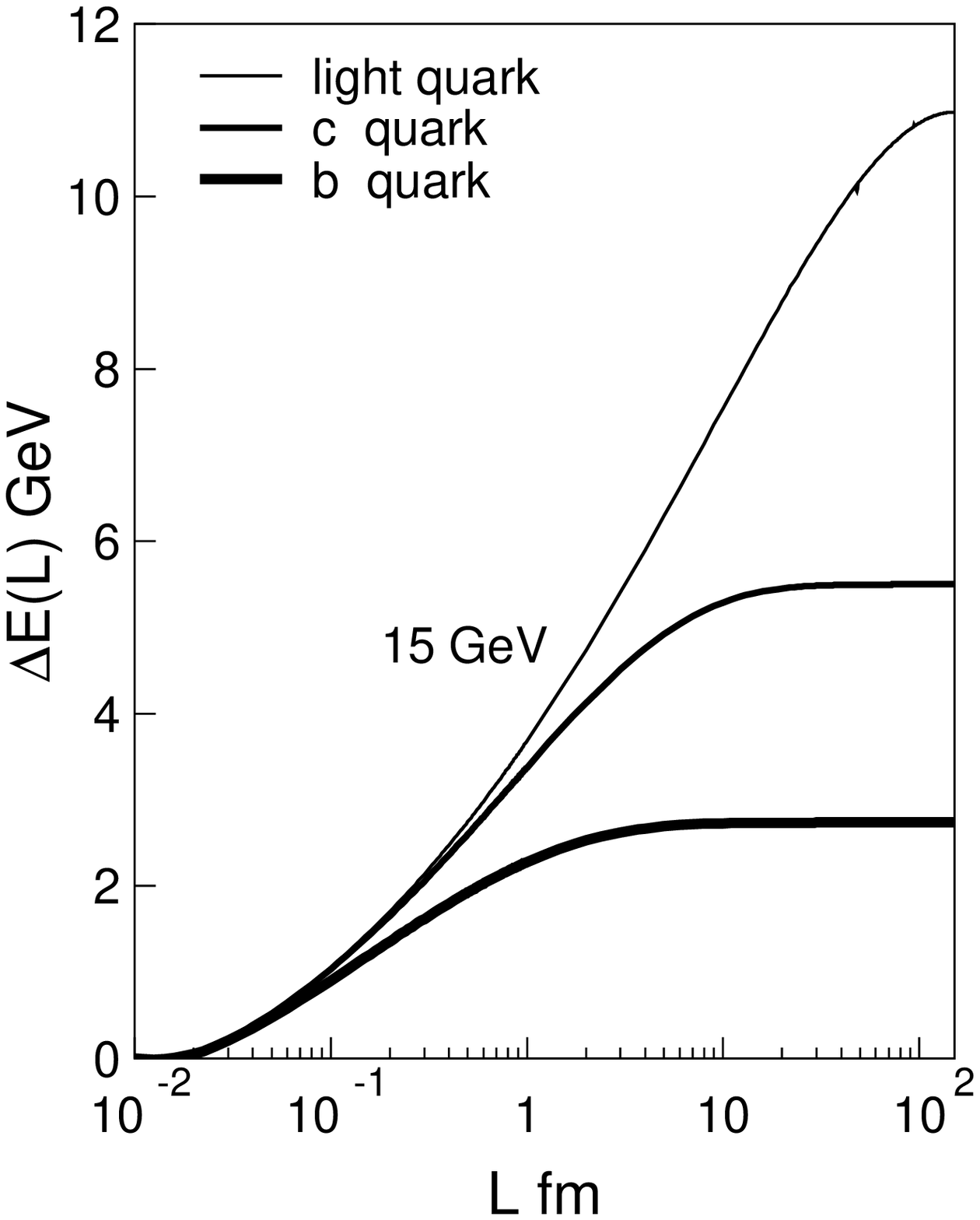}\hspace{10mm}
 \includegraphics[height=7cm]{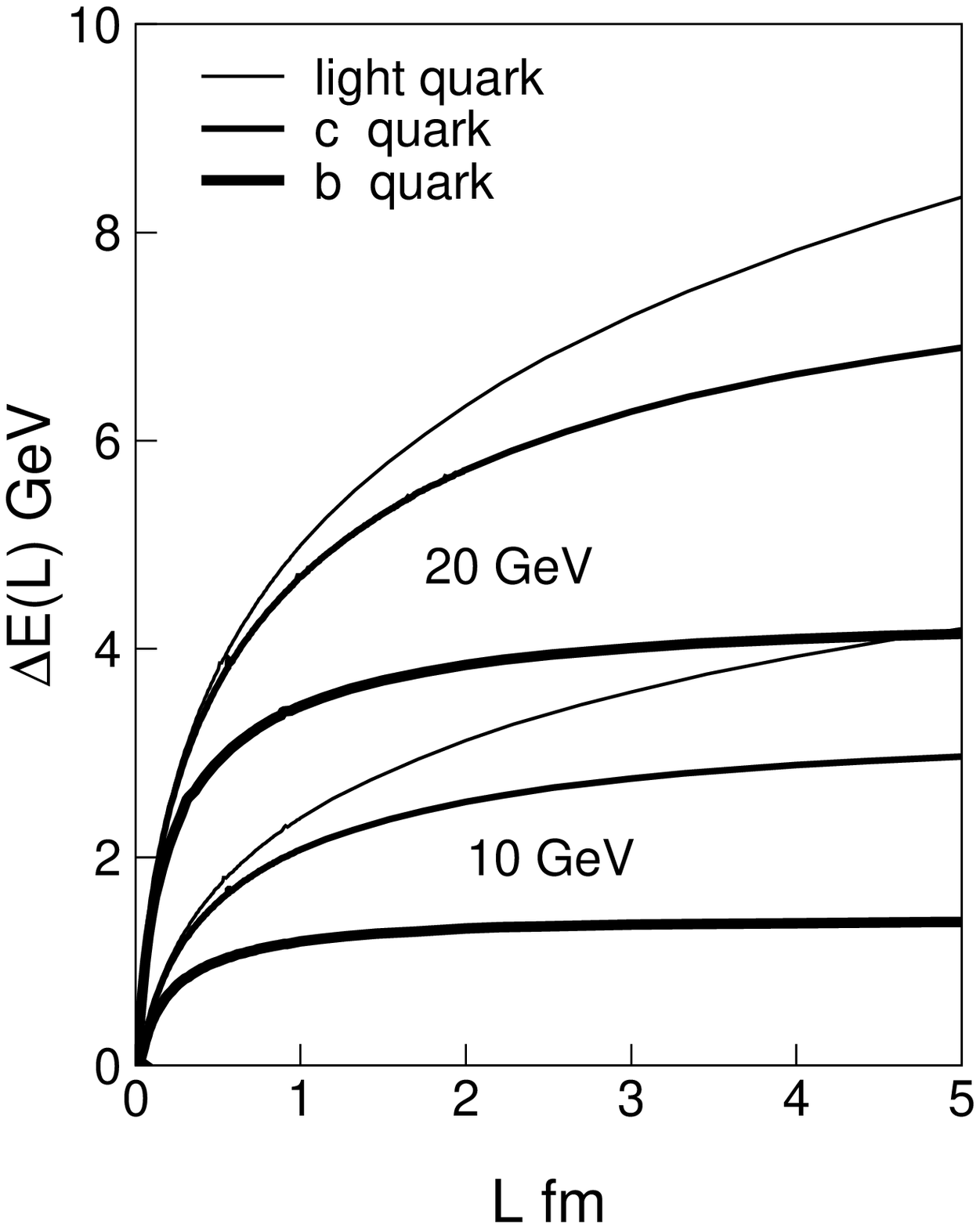}
\caption{\label{l-dep} Upper panel: vacuum energy loss by light ($m_q=0$), charm ($m_c=1.5\GeV$) and bottom ($m_b=4.5\GeV$) quarks with $E=15\GeV$ as function of path length. Lower panel: the same, but for energies $10$ (three upper curves) and $20\GeV$ (three bottom curves), and zoomed in at short path lengths.
}
 \end{figure}

The numerical results demonstrating this behavior are depicted in
Fig.~\ref{l-dep}.
One can see that a substantial difference between radiation of energy by the charm and light quarks
onsets at rather long distances, above $10\fm$, while  within several fermi the difference is insignificant. The $b$-quark radiation is suppressed already at rather short distances.

Apparently, this concerns not only the vacuum, but also medium induced radiation, which is softer and is even more affected by the vacuum dead cone.   As far as the field of the quark is not regenerated yet, it is lacking for induced radiation as well.

Although at sufficiently long distances the field of the quark is
restored and a heavy quark starts radiating much less than a light one,
this difference does not show up in heavy ion collisions observed at
RHIC. Indeed, data show that high-$p_T$ hadrons produced  in central
gold-gold collisions at $\sqrt{s}=200\GeV$ are about five times
suppressed \cite{phenix}. This means that not the whole volume of the
produced quark-gluon medium, but only a rather thin outer
layer of the medium contributes. One can estimate the thickness of
this layer assuming it to be transparent for partons, but the
deeper layers to be fully opaque. The amount of produced high-$p_T$
partons at impact parameter $b$ is proportional to
$T_A^2(b)=4\rho_0(R_A^2-b^2)$, where $R_A$ and $T_A(b)$ are the
radius and thickness function for a nucleus with constant density
$\rho_0$. The fraction of the number of hard collisions, which occur
in the outer layer of thickness $\Delta\ll R_A$, is $R_{AA}(\Delta)= 10\,\Delta^2/R_A^2$. 
For $R_A=6.5\fm$ and $R_{AA}=0.2$ we get $\Delta=0.9\fm$. This rough
estimate shows that the actual distance covered by a high-$p_T$
parton inside the produced dense medium is short and allows 
to neglect the small difference between the rates of energy
loss (either vacuum, or induced) of charm and light quarks.
Therefore $R_{AA}$ should be also independent of quarks mass. 

At the same time, Fig.~\ref{l-dep} demonstrates that energy loss of $b$-quarks is
significantly reduced already at very short distances, less than fermi, 
so beauty should be much less suppressed in nuclear collisions. 
Only at about an order of magnitude larger $p_T$
$b$-quarks will radiate and get suppressed
as much as other flavors.

Concluding, a highly virtual quark originated from a hard reaction
with its color field stripped off develops a vacuum dead-cone which
is controlled by its virtuality. While the virtuality is high,
$Q^2(L)\gg m_q^2$, this cone is wider than the one related to the
quark mass. Therefore, either vacuum, or medium-induced energy loss
proceed with the rates independent of the quark mass. Only at longer
distances, when the quark virtuality decreases down to the scale of
the heavy quark mass, its radiation gets reduced compared to the
light quark. For a charm quark this regime is reached at a rather
long distance of several fermi, outside the dense medium created in
heavy ion collisions. This explains the flavor independent
suppression observed at large $p_T$ at RHIC. We expect, however, that open beauty
should be much less suppressed than other hadrons within the same
range of $p_T$, but will reach the universal magnitude of
suppression at much larger $p_T$. Notice that the above
consideration does not depend on the collisions energy $\sqrt{s}$,
but only on $p_T$.

\begin{acknowledgments}

We are thankful to Yuri Dokshitzer for encouraging  discussions. This work was supported in part
by Fondecyt (Chile) grants 1090236, 1090291 and 1100287, and by DFG
(Germany) grant PI182/3-1.

\end{acknowledgments}

\end{document}